\documentclass[aps,prl,twocolumn,groupedaddress,showpacs]{revtex4}
\begin{document}
\title{Information capacity of optical fiber channels with zero average dispersion}
\author{K. S. Turitsyn$^{1}$, S. A. Derevyanko$^{2,3}$, I. V. Yurkevich$^{4}$, S. K.
Turitsyn$^{2}$}

\affiliation{$^{1}$Landau Inst. for
Theor. Physics, Moscow, Kosygina 2, 117940, Russia \\
$^{2}$Photonics Research Group, Aston University, Birmingham B4 7ET, UK\\
$^{3}$Institute for Radiophysics and Electronics, Kharkov 61085,
Ukraine \\$^{4}$The Birmingham University, Birmingham B15 2TT, UK}

\begin{abstract}
We study the statistics of optical data transmission in a noisy nonlinear
fiber channel with a weak dispersion management and zero average dispersion.
Applying path integral methods we have found exactly the probability density
functions of channel output both for a non-linear noisy channel and for a
linear channel with additive and multiplicative noise. We have obtained
analytically a lower bound estimate for the Shannon capacity of considered
nonlinear fiber channel.
\end{abstract}

\maketitle


\textit{Introduction} The classical theorem of information theory
\cite {Shannon} states that the capacity of a power-constrained
transmission in an additive Gaussian noise channel growths
logarithmically with increase of the signal to noise ratio (SNR).
Thus, an improvement of the capacity (maximum average information
per symbol that can be transmitted through the channel) in such
systems can be achieved by increase of the signal power assuming
that the noise level is not affected. The Gaussian statistics of
noise is a fundamental assumption in derivation of this widely
known Shannon's result. Properties and applications of bandlimited
linear channels with additive white Gaussian noise (AWGN) form a
foundation of modern information theory. It should be emphasized
that the AWGN linear channel model is not just a simple
mathematical construction, but is applied directly to many
practical problems such as, for instance, deep-space
communication. However, in some applications, nonlinear response
of a transmission medium must be taken into account. Evidently,
properties of nonlinear information channel can be significantly
different from that for AWGN models. Interaction of noise and
signal in nonlinear transmission channel can result in
non-Gaussian statistics of received signals. The theory of
non-Gaussian information channels though being an evident
challenge for many decades is not yet well established compared to
the success of AWGN models. Studies in this fundamental research
area are additionally motivated by practical technological
achievements and growing demand for efficient high-speed, high
quality communications. Recent progress in fiber optics attracts
much fresh interest to the information theory of non-Gaussian
nonlinear communication channels \cite{Mitra}-\cite{Mitra2}.
Optical fiber waveguides made of silica present low loss,
ultra-high capacity, cost-efficient transmission media with many
attractive features. Using optical amplifiers to recover signal
power simultaneously at many carrier frequencies (channels) within
the fiber bandwidth it is possible to transmit optical information
data over thousands of kilometres. It is already well recognized,
however, that the nonlinear response of the transmission medium
(Kerr effect nonlinearity) plays a crucial role in limiting the
aggregate capacity of optical fiber systems. Accumulation of
nonlinear interactions with propagation along the transmission
line makes fiber information channels essentially nonlinear.
Evidently, nonlinear impairments (or in other words, a level of
signal corruption due to nonlinearity) depend on the signal power.
Therefore, in nonlinear channels an increase of the signal power
would not necessarily improve the system capacity. Recently, in
their pioneering work Mitra and Stark suggested that from the
information theory perspective under certain conditions one can
treat essentially nonlinear noisy channels as linear ones with
effective multiplicative noise \cite{Mitra}. Applying this idea to
multi-channel optical fiber transmission systems they derived a
heuristic linear model with multiplicative noise that presumably
approximates some features of the original nonlinear channel.
Though a connection between statistical properties of such an
effective ''nonlinear noise'' and system/signal characteristics is
still a subject of further research and justification, this
intuitive approach outlines a possible way to treat nonlinear
transmission channels. In order to compute the Shannon capacity it
is necessary to make one more step beyond determination of a
conditional probability. Namely, one has to find the optimal input
signal statistics
(that is even more complicated functional problem). The channel capacity $%
\mathcal{C}$ defined by Shannon is a maximum of the following functional
(called mutual information) with respect to the statistics of input signal $X
$, given by distribution function $p(X)$:
\begin{equation}
\mathcal{C}=\max_{p(X)}\int \mathrm{\mathcal{D}X\mathcal{D}Y}P(X,Y)\log _{2}%
\frac{P(X,Y)}{P_{out}(Y)p(X)}.  \nonumber
\end{equation}
Here $P(X,Y)=P(Y|X)\,p(X)$ is the joint distribution function of input $X$
and output $Y$; $P_{out}(Y)=\int \mathcal{D}\mathrm{X}P(Y|X)p(X),$ and all
specific properties of a communication channel are given by the conditional
probability $P(Y|X).$ To the best of our knowledge the only case for which
there exists an explicit analytical solution of the corresponding functional
optimization problem is when the joint distribution of input and output
signals are Gaussian. In this case the Shannon capacity can be explicitly
expressed through an input-output pair correlation matrix introduced by
Pinsker \cite{Pinsker}. Difficulties in the analysis of non-Gaussian
nonlinear channels to some extent are caused by a relatively limited number
of appropriate mathematical methods. Therefore, to practically estimate the
capacity of the nonlinear fiber channel most authors \cite{Tang1}-\cite
{Mitra2} apply the Pinsker formula that for the Gaussian noise coincides
with Shannon's definition, but as a matter of fact gives only the lower
bound on the capacity \cite{Mitra}. An interesting open problem is
capability of the Pinsker formula to mimic the true behavior of the capacity
of nonlinear information channels especially in the case of large input
signal power. Computation of Shannon capacity for any realistic optical
amplifier transmission system is a very complicated problem which is
unlikely to be solved analytically. Therefore, it is of crucial importance
for further progress in this area to find basic simplified models of fiber
nonlinear channels that can be treated analytically. Such solvable models
can provide guidance to analysis of much more complicated general problems
in the information theory of nonlinear fiber channels.

In this Letter we present a theoretical analysis of \ a physical model which
describes the transmission of light signals in a noisy nonlinear fiber
channel with zero average dispersion. To examine the similarity and
difference between the effects of nonlinearity and multiplicative noise, in
parallel, we study a linear model of the channel with both additive and
multiplicative noise. We calculate analytically the probability density
function (PDF) of the channel output for both models. Using our derived
conditional probabilities we analyze the capacity of corresponding
transmission systems. We compare here two approaches to the estimation of
system capacity: first, based on Pinsker's formula for input-output
correlation matrix and, second, directly applying Shannon's definition of
the capacity.

The average propagation of a complex light envelope $E(z,t)$ in a noisy
optical fibre line with the so-called weak dispersion management (see for
details e.g. \cite{Mecozzi}, \cite{Tur1}) in the main order is described by
the stochastic nonlinear Schr\"{o}dinger equation$:$
\begin{equation}
\frac{\partial E}{\partial z}=i\,\frac{<d>}{2}\frac{\partial ^{2}E}{\partial
t^{2}}+i|E|^{2}E+n.  \label{nse}
\end{equation}
Here $n(z,t)$ is an additive complex white noise with zero mean and
correlation function (see for notations \cite{Mecozzi})
\begin{equation}
<n(z,t)n^{\ast }(z^{\prime },t^{\prime })>=<n_{0}>\delta (z-z^{\prime
})\delta (t-t^{\prime }).
\end{equation}
In the present Letter we restrict consideration to the case of weakly
dispersion-managed fiber systems with  zero average dispersion $<d>=0$. The
propagation equation then is effectively reduced to the Langevin equation
for the regularized field $u(z)\equiv E(z,0)$ with the regularized noise $%
\eta (z)\equiv n(z,0)$.
\begin{equation}
\frac{du}{dz}-i|u|^{2}\,u=\eta ,\qquad u(z=0)=u_{0},\qquad   \label{lang}
\end{equation}
Here $\eta (z)$ is a white noise with zero mean and correlation function $%
<\eta (z)\eta ^{\ast }(z^{\prime })>=D\delta (z-z^{\prime })$, where $%
D=2W<n_{0}>$ is the regularized noise intensity. To restore the capacity for
a bandwidth limited signal one has simply to multiply all the corresponding
results by the channel bandwidth $W$.

\textit{Calculation of a conditional probability} Some statistical
properties of system (\ref{lang}) including higher-order momenta have been
studied by Mecozzi \cite{Mec2}. However, the method used in \cite{Mec2} did
not permit explicit computation of the PDF which is required in the analysis
of system capacity. Therefore, to calculate the conditional probability $%
P(u,z|u_{0})$ we apply here the so-called Martin-Siggia-Rose formalism \cite
{Zinn-Justin} that presents the conditional PDF of the output as the
following functional integral:
\begin{equation}
P(u,z|u_{0})=\int\limits_{q(0)=u_{0}}^{q(z)=u}\mathcal{D}q\,e^{-\int%
\limits_{0}^{z}\mathrm{d}z^{\prime }\mathcal{L}\left[ q(z^{\prime })\right]
},  \label{cond}
\end{equation}
where the effective Lagrangian is defined as $\mathcal{L}[q]=(2D)^{-1}|q^{%
\prime }-i|q|^{2}\,q|^{2}.$ Integral (\ref{cond}) can be calculated
analytically. The substitution $q(z)={\tilde{q}}(z)\,\exp [i\int_{0}^{z}%
\mathrm{d}z^{\prime }|{\tilde{q}}(z^{\prime })|^{2}]$ brings Lagrangian to
its free form. The Jacobian of this transform is unity and in the new
variables the integral becomes Gaussian. After simple straightforward
algebra it can be reduced to
\begin{equation}
P(u,z|u_{0})=\sum\limits_{m=-\infty }^{+\infty }e^{im\phi }\int \frac{%
\mathrm{d}\phi ^{\prime }}{2\pi }\,e^{-im\phi ^{\prime }}{P}^{^{\prime
}}(r,\phi ^{\prime },z|r_{0},\phi _{0})
\end{equation}
where the auxiliary ``partition function'' is
\begin{equation}
{P}^{^{\prime }}(r,\phi ^{\prime },z|r_{0},\phi _{0})\equiv
\int\limits_{q(0)=r_{0}\,e^{i\phi _{0}}}^{q(z)=r\,e^{i\phi ^{\prime }}}%
\mathcal{D}q\,e^{-\int\limits_{0}^{z}\mathrm{d}z^{\prime }\{i\,m\,|q|^{2}+%
\frac{1}{2D}|q^{\prime }|^{2}\}}
\end{equation}
(here $u=re^{i\phi }$, $u_{0}=r_{0}e^{i\phi _{0}}$). The effective
action decomposes into sum of the classical part and a fluctuating
part that does not depend on the limits. The fluctuating field is
calculated by expanding over the complete set of eigenfunctions of
the operator $-\partial _{z}^{2}+k_{m}^{2}$ satisfying zero
boundary conditions at $z^{\prime }=0$ and $z^{\prime }=z.$
Omitting details of these operations we present a final expression
for the conditional probability of our nonlinear channel:
\[
P(u,z|u_{0})=\frac{1}{2\pi }\,\sum\limits_{m=-\infty }^{+\infty
}\,e^{im(\phi -\phi _{0})}\,P_{m}(r,z|r_{0})
\]
\begin{equation}
 =\frac{1}{2\pi
D}\,\sum\limits_{m=-\infty }^{+\infty }\,\frac{e^{im(\phi
-\phi _{0})}\,e^{-\frac{r^{2}+r_{0}^{2}}{2D}k_{m}\coth k_{m}z}}{\sinh k_{m}z}%
\,\,k_{m}\,\mathrm{I}_{|m|}(q_{m})\,,  \label{pdf1}
\end{equation}
here $q_{m}=k_{m}r\,r_{0}/(D\sinh (k_{m}z))$, $k_{m}=\sqrt{2imD}$
and $\mathrm{I}_{|m|}$ is the modified Bessel function.

Next we establish an analogy between the considered nonlinear channel (NLCH)
and a linear channel with multiplicative noise (LMNCH):
\begin{eqnarray}
u^{\prime }-iv\,u &=&\eta ,\quad u(z=0)=u_{0},  \label{fake} \\
&<&\eta ^{\ast }(z^{\prime })\eta (z)>=D\delta (z-z^{\prime }), \\
&<&v(z)v(z^{\prime })>=D^{\prime }\delta (z-z^{\prime })
\end{eqnarray}
Applying a similar procedure to above we derive the conditional probability
function of the form Eq.(\ref{pdf1}) with replacement $P_{m}\rightarrow {%
\tilde{P}}_{m}$ where
\begin{equation}
{\tilde{P}}_{m}(r,z|r_{0})=\frac{1}{Dz}\,e^{-m^{2}D^{\prime }z/2}\,\mathrm{I}%
_{|m|}\left( \frac{rr_{0}}{Dz}\right) \,e^{-\frac{r^{2}+r_{0}^{2}}{2Dz}}.
\label{pdf2}
\end{equation}
Note that if the information is transmited using only signal power (the
so-called intensity modulatuion - direct detection systems) $r=|u|$ the
conditional probability takes the form (after integration in polar
coordinates $(r,\phi )$ over phase $\phi $: $\int d\phi
P(u,z|u_{0})=P_{0}(r,z|r_{0}))$:
\begin{equation}
P_{0}(r,z|r_{0})=\widetilde{P}_{0}(r,z|r_{0})=\frac{1}{\,Dz}\,\,\mathrm{I}%
_{0}\left( \frac{r\,r_{0}}{zD}\right) \,e^{-\frac{r^{2}+r_{0}^{2}}{2\,z\,D}}.
\label{p-r}
\end{equation}

\bigskip Note that in both cases (nonlinear and effective multiplicative
noise channels) formulae (\ref{pdf1}) and (\ref{pdf2}) yield the same result.

\textit{Channel capacity} First we revise the procedure commonly used in the
recent literature for the channel capacity estimation. We demonstrate here
that the consideration based on pair correlation functions \cite{Pinsker}
can lead to results very different from the Shannon capacity and, therefore,
should be used with caution. Some authors \cite{Tang1,Tang2} instead of
using the original Shannon definition calculate capacity by exploiting a
simpler Pinsker formula based on a complex self-conjugate \textit{%
input-output correlation matrix }$C_{\alpha \beta }$:
\begin{equation}
\mathcal{C}_{G}\equiv \log _{2}\frac{\mathrm{Det}\,\mathrm{diag}(C_{\alpha
\beta })}{\mathrm{Det}\,C_{\alpha \beta }},\;\;C_{\alpha \beta }\equiv
<u_{\alpha }u_{\beta }^{\ast }>  \label{gauss-cap}
\end{equation}
Here indices $\alpha ,\beta =input,output$; and brackets stand for the
average over noise ($\eta $ for non-linear problem and $\eta $ and $v$ for
double noise model) and over statistics of the input signal $u_{0}$ (which
is assumed to be Gaussian). Defined in this way the Gaussian capacity $%
\mathcal{C}_{G}$ coincides with the Shannon capacity for the case of
Gaussian joint input-output distributions which corresponds to the linear
channel with additive noise \cite{Pinsker}. For nonlinear channels or
channels with multiplicative noise the Gaussian capacity (\ref{gauss-cap})
represents the lower estimate for the true Shannon capacity $\mathcal{C}$
(see \cite{Mitra}).

We start from the calculation of the correlation matrix. To perform noise
averages we use either PDF (\ref{pdf1}) or (\ref{pdf2}). It is easy to find
that $C_{in,in}=\left\langle \left| u_{0}\right| ^{2}\right\rangle \equiv
S\,,\;C_{out,out}=\left\langle \left| u(z)\right| ^{2}\right\rangle =(S+N),$
$N\equiv 2D\,z$ regardless the model. However, the cross-correlations $%
C_{in,out}=\left\langle u_{0}\,u^{\ast }(z)\right\rangle $ are different
\begin{equation}
C_{in,out}=\left\{
\begin{array}{c}
\frac{S\,\mathrm{sech}^{2}k_{1}z}{\left( 1+(S/N)k_{1}z\tanh k_{1}z\right)
^{2}},\;\mathrm{NLCH} \\[6mm]
S\,e^{-D^{\prime }z/2},\;\mathrm{LMNCH}
\end{array}
\right\}
\end{equation}
where $k_{1}=\sqrt{2iD}$. Note that SNR $=S/N=s$ changes only due to
variation of $S$, while $N=2Dz$ is fixed as we consider here a fixed
transmission distance. Substitution of the correlation matrix into the
definition Eq.(\ref{gauss-cap}) yields the final result
\begin{equation}
\mathcal{C}_{G}=\log _{2}\left[ 1+\frac{s}{\left( 1+s\right)
\,a(z)|1+s\,b(z))|^{4}-s}\right] ,  \label{res}
\end{equation}
where
\begin{equation}
a(z)=\left\{
\begin{array}{cc}
|\cosh k_{1}z|^{4}, & \mathrm{NLCH} \\
e^{D^{\prime }z}, & \mathrm{LMNCH}
\end{array}
\right.
\end{equation}
\begin{equation}
b(z)=\left\{
\begin{array}{cc}
k_{1}z\tanh k_{1}z, & \mathrm{NLCH} \\
0, & \mathrm{LMNCH}
\end{array}
\right.
\end{equation}
It is seen from Eq. (\ref{res}) that with increase of SNR $\mathcal{C}_{G}$
decays to zero for the case of the nonlinear channel (simular to conclusions
made in \cite{Tang1,Tang2}) and tends to a constant for the case of the
multiplicative noise channel. However, below we will show that in both cases
the true Shannon capacity $\mathcal{C}$ is unbounded and grows
logarithmically with increase of $S/N$ similar to the linear channel.

\textit{Direct estimate of the Shannon capacity} Following Shannon \cite
{Shannon} we consider now the channel capacity $\mathcal{C}$ defined as a
maximum of the mutual information with respect to the statistics of input, $%
u_{0}$, given by distribution function $p(u_{0})$ under the fixed average
input power $S$:
\begin{equation}
\mathcal{C}=\max_{p(u_{0})}\int \mathrm{d}^{2}u\mathrm{d}^{2}u_{0}P(u,u_{0})%
\log _{2}\frac{P(u,u_{0})}{P_{out}(u)p(u_{0})}.  \nonumber
\end{equation}
The conditional probability $P(u|u_{0})$ connecting output and input
probabilities: $P_{out}(u)=\int \mathrm{d}^{2}u_{0}P(u|u_{0})p(u_{0})$ is
given either by (\ref{pdf1}) or (\ref{pdf2}). Note that the Shannon
definition allows one to obtain directly an estimate of capacity. Any
arbitrary trial distribution $p(u_{0})$ provides for a certain low boundary
estimate of the capacity $\mathcal{C}$. The closer a trial function is to
the optimal distribution of $p(u_{0})$ the better is our approximation of
the true capacity. Applying the so-called Klein inequality for two arbitrary
probability distribution functions $P$ and $\mathcal{P}$
\begin{equation}
\int \mathrm{d}^{2}u\mathrm{d}^{2}u_{0}P(u,u_{0})\log _{2}\frac{P(u,u_{0})}{%
\mathcal{P}(u,u_{0})}\geq 0
\end{equation}
we obtain the following chain of inequalities:
\begin{eqnarray}
\mathcal{C} &\geq &\int \mathrm{d}^{2}u\mathrm{d}^{2}u_{0}\,P(u,u_{0})\,\log
_{2}\frac{P(u,u_{0})}{P_{out}(u)p(u_{0})}  \nonumber \\
&\geq &\int \mathrm{d}^{2}u\mathrm{d}^{2}u_{0}\,P(u,u_{0})\,\log _{2}\frac{%
\mathcal{P}(u,u_{0})}{P_{out}(u)p(u_{0})}  \label{chain}
\end{eqnarray}
where $\mathcal{P}$ is an arbitrary PDF (by this we mean that it is
non-negative and normalized) and $p(u_{0})$ is an arbitrary (not optimal)
initial signal distribution. Next we exploit an arbitrariness of $\mathcal{P}
$ and $p(u_{0})$ in (\ref{chain}) by chosing $p(u_{0})=(2\pi
)^{-1}p(r_{0}),\quad \mathcal{P}(u,u_{0})=(2\pi
)^{-2}P_{0}(r|r_{0})\,p(r_{0}).$ Here we assumed that both an input
distribution $p(u_{0})$ and $\mathcal{P}$ are phase independent and $%
P_{0}(r|r_{0})$ is the radial conditional probability given by Eq.(\ref{p-r}%
). Substitution of these trial functions into inequality Eq.(\ref{chain})
brings it to the form
\begin{equation}
\mathcal{C}\geq \int \mathrm{d}r\mathrm{d}r_{0}\,r\mathrm{\,}%
r_{0}P_{0}(r|r_{0})\,p(r_{0})\log _{2}\frac{P_{0}(r|r_{0})}{\int \mathrm{d}%
r^{\prime }r^{\prime }P_{0}(r|r^{\prime })p(r^{\prime })}.  \label{ineq}
\end{equation}
Evaluation of the r.h.s. of this inequality for any trial function leads to
an estimate of a lower bound for the Shannon capacity. Substituting $%
P(r,z|r_{0})$ from (\ref{p-r}), and considering a Gaussian trial function
for input statistics $p(r_{0})=(2/S)\exp (-r_{0}^{2}/S)$, after simple
algebra we obtain:
\begin{eqnarray}
\mathcal{C} &\geq &\mathcal{C}_{0}(s)=\ln (1+s)-2s+F_{1}(s) \\
F_{1}(s) &=&s^{-1}\,\int_{0}^{\infty }\mathrm{d}x\,x\mathrm{K}_{0}(x\sqrt{%
1+s^{-1}})\mathrm{I}_{0}(x)\ln \mathrm{I}_{0}(x)  \nonumber  \label{cap}
\end{eqnarray}
where $\mathrm{I}_{0}$ and $\mathrm{K}_{0}$ are modified Bessel functions
and $s=S/N$ is the SNR. Then the main contribution from the integral $%
F_{1}(s)$ to the asymptotic behavior of $\mathcal{C}_{0}(s)$ for large $s$
comes from the region $x\gg 1.$ Using the asymptotic expansion of modified
Bessel functions we get
\[
\mathcal{C}\geq \frac{1}{2}\ln s+O(1).
\]
This proves that $\mathcal{C}_{0}$ and hence the Shannon capacity $\mathcal{C%
}$ are both unbounded as $S/N\rightarrow \infty $.

Our result in particular shows that a naive straightforward application of
the Pinsker formula for evaluation of the capacity of a nonlinear channel
as, for instance, in \cite{Tang1,Tang2} can lead to wrong conclusions
regarding the asymptotic behavior of the capacity with $S/N\rightarrow
\infty $. Note that for the specific problem considered here it is possible
to modify the definition of $C_{\alpha \beta }$ to obtain correct
asymptotics for capacity using input-output correlation matrix. Indeed,
calculation of $\mathcal{C}_{G}\mathit{\ }$constructed with correlators $%
\left\langle r_{0}^{2}\right\rangle =S,\quad \left\langle r\right\rangle
=S+N,\quad \left\langle r\,r_{0}\right\rangle =S\,F_{2}(s),$ $%
F_{2}(s)=(2s^{2})^{-1}\int_{0}^{\infty }\mathrm{d}x\,x^{2}\,\mathrm{I}%
_{0}(x)\,\mathrm{K}_{0}(x\sqrt{1+s^{-1}}),$ leads to $\mathcal{C}_{G}=\ln
(1+s)/(1+s\left[ 1-F_{2}(s)\right] ).$ Taking into account that $%
F_{2}(s\rightarrow \infty )\rightarrow 1$ one can see that it gives the
correct asymptotic behavior for capacity. Unfortunately, there is no general
recipe for choosing the correct correlators in the Pinsker formula.

\textit{Discussion and conclusions }We have examined the statistics of
optical data transmission in a noisy nonlinear fiber channel with a weak
dispersion management and zero average dispersion. We have also studied
similarity and difference between effects of nonlinearity and multiplicative
noise, considering in parallel a linear channel with multiplicative (and
additive) noise. Using analytically calculated conditional PDF we analyzed
the Shannon transmission capacity for both models. We did manage to find
analytically a lower bound estimate for the Shannon capacity of the
nonlinear fiber channel considered here. We revise the Pinsker formula which
has been used without justification in some recent works and show that the
Gaussian capacity defined through the pair correlation functions should be
used with caution in the case of nonlinear transmission channels. To
incorporate the optimization procedure inherent for the Shannon definition
one needs to elaborate all possible correlators and find those which are
essential, i. e. are much greater than others. Those correlators may then be
used in the Pinsker formula to provide a simple and tractable expression for
the channel capacity. That would not be necessary if the Shannon definition
could be worked out. Unfortunately, it is hardly the case for any more or
less practical problem of interest. Another important result of our analysis
is that nonlinearity and multiplicative noise do not necessarily degrade
input-output correlations in the same way. Therefore, relating the nonlinear
problem to a linear one with multiplicative noise has to be carefully
justified for each specific transmisison system model.
\begin{acknowledgments}
This work was supported by INTAS Young Scientist Fellowship No YS
2002 - 165 (S.D.) and by the Liverhulme Trust project A/20010049
(S.D., S.K.T.). I.V.Y. gratefully acknowledges support by  the
Leverhulme Trust under the contract F/94/BY and by the EPSRC grant
GR/R95432.
\end{acknowledgments}

\end{document}